\documentclass[aps,pra,floatfix,twocolumn,showpacs]{revtex4}
\usepackage{amsmath}
\begin{document}
\title{Calculation of isotope shifts for cesium and francium}
\author{V. A. Dzuba}
\email{V.Dzuba@unsw.edu.au}
\affiliation{School of Physics, University of New South Wales, Sydney 2052,
Australia}

\author{W. R. Johnson}
\email{johnson@nd.edu}
\homepage{www.nd.edu/~johnson}
\affiliation{
Department of Physics, 225 Nieuwland Science Hall\\
University of Notre Dame, Notre Dame, IN 46566}

\author{M. S. Safronova}
\email{msafrono@physics.udel.edu}
\affiliation{
Department of Physics and Astronomy,
University of Delaware, Newark, Delaware 19716}

\date{\today}

\begin{abstract}

We perform {\em ab initio} calculations of isotope shifts for 
isotopes of cesium (from $A$=123 to $A$=137) and francium
(from $A$=207 to $A$=228). These calculations start from the relativistic
Hartree-Fock method and make use of several techniques to include
correlations. The field (volume) isotope shift is calculated by means
of an all-order correlation potential method and within the
singles-doubles partial triples linearized coupled-cluster approach.
Many-body perturbation theory in two different formulations is
used to calculate the specific mass shift. We discuss the strong points and
shortcomings of the different approaches and implications for parity
nonconservation in atoms. Changes in nuclear charge radii
are found by comparing the present calculations with experiment.

\end{abstract}



\maketitle

\section{introduction}

Accurate calculations of isotope shifts for many-electron atoms are needed
to address a number of pressing problems. 
Although possible changes of isotope abundances 
in the early universe are important systematic effects in the study of variation of
fundamental constants \cite{alpha1,alpha2,alpha3}, isotope 
shift data are unavailable for most of the spectral lines of interest. 
Furthermore, studies of parity nonconservation (PNC) in atoms
have come to the point where more information on nuclear structure is needed. 
The analysis of the most precise measurement of PNC in cesium reveals that at
the present level of accuracy (0.35\% for experiment, 0.5\% for theory,
see, e.g. \cite{ginges}) there is
perfect agreement with the standard model of the electroweak interactions. 
To search for new physics beyond the
standard model in low-energy physics, one needs to improve the accuracy of the
analysis. While improving the accuracy of measurements is probably feasible, 
improving the accuracy of calculations is problematic. To avoid problems
with the accuracy of calculations, 
\citet{Khriplovich}  suggested that PNC be measured in a chain
of isotopes. The electronic structure factor would be canceled in the ratio
for different isotopes. However, at the level of accuracy
needed to search for new physics, the cancellation is not sufficiently complete.
One must account for the change in electronic structure caused by the
change in nuclear charge radius. The change in neutron distribution presents a
separate  problem \cite{Fortson} that will not be discussed further in this work, 
since the isotope shift is not
sensitive to the changes in the neutron distribution. 
The change in the nuclear charge radius 
can be obtained by comparing experimental and theoretical values of isotope shifts.
Moreover, calculations of the isotope shift provide tests of atomic wave functions at
short distances, thereby providing another way of evaluating the accuracy of atomic
PNC calculations. 

We have chosen cesium and francium for our analysis since both atoms are being
used or planed to be used in atomic PNC measurements. Also, both atoms have relatively 
simple electronic structure with one external electron above closed shells, 
making it easier to study different theoretical approaches. Accurate experimental values
of isotope shifts for a range of isotopes of both atoms are available.

Isotope shifts of cesium, francium, and thallium have been studied theoretically
by A.-M. M{\aa}rtensson-Pendrill {\em et al.} \cite{ann-pnc,ann-cs-tl,ann-fr}.
In this paper, we present several different approaches aimed at improving
the accuracy of 
isotope shift calculations and the corresponding analysis of changes in 
nuclear charge radii. 
We demonstrate that, owing to the
extremely poor convergence of many-body perturbation theory, all-order
techniques are needed to obtain reliable results.

In heavy atoms, the isotope shift is dominated by the field (or volume) shift (FS), which is the
shift in energy caused by the change in the nuclear charge radius. In light atoms, the isotope
shift is dominated by the mass shift, which is the change in energy due to differences 
in isotopic masses. 
Both of these effects are important for cesium. 
The field shift strongly dominates for francium. The FS is easier to calculate owing
to the simpler form of the corresponding  operator. We present two different all-order 
methods, both of which
use the finite-field approach to reduce the calculation of the FS to a calculation 
of energy. The first method is the all-order correlation potential method and the second is the
singles-doubles, partial triples, linearized coupled-cluster method. 
In order to reduce the calculation of the FS to a calculation of an energy, we incorporate the 
rescaled operator of the derivative of the nuclear potential with respect to the nuclear radius
into the Hamiltonian at each stage of the calculation.
The isotope shift is calculated as the derivative of the energy with respect to the scaling
parameter. Both methods give very close results. Calculations of energies and hyperfine structures 
(hfs) are used as additional tests of the accuracy of the calculations. 
We believe that the uncertainty in our calculation of the field shift does not exceed 1\%.

For reasons explained below, the same methods cannot be used at the present time for the
calculation of the mass shift. 
Therefore, we use two different approaches there. The first approach is a
third-order perturbation theory calculation (with certain classes of terms included to all orders).
The second method which we use in this work is a combination of perturbation theory and the
finite-field approach in which only core polarization diagrams are included to all
orders. While such a treatment of the specific mass shift allowed us to obtain
reasonably good results suitable for the present work, we
stress the need to develop a complete all-order technique for the calculation of the mass shift,
similar to that used here for the field shift.

We extract values of the change in nuclear charge radius 
for different isotopes of cesium and francium by comparing 
our calculations with available
experimental data and we discuss the implications of
this work for PNC in atoms.

\section{Method of Calculations}
\label{method}

The shift of transition frequency of isotope $A'$ compared to isotope $A$ can be written as
\begin{equation}
        \delta\nu_{AA'} = (K_\text{NMS} + K_\text{SMS})(\frac{1}{A}-\frac{1}{A'})+
	F\delta \langle r^2 \rangle ^{AA'},
\label{IS}
\end{equation}
where $A$ and $A'$ are mass numbers of the two isotopes and $\langle r^2 \rangle$ 
is the mean square nuclear radius.
 The first term on the right represents the
mass shift and the second term represents the field shift.
The mass shift consists of two parts: a normal mass shift (NMS) and a specific mass shift (SMS).
The normal mass shift constant is expressed in terms of the experimental frequency by
\begin{equation}
        K_\text{NMS} = \frac{\nu_\text{exp}}{1822.888},
\label{NMS}
\end{equation}
while SMS and FS constants $K_\text{SMS}$ and $F$ are the subjects of the present calculations.

\subsection{Field shift}

We start with field shift calculations because they are easier to carry out and because they
play more important roles in our analysis owing to their implication for PNC in
atoms and their dominance for the francium isotope shift.

We use the following form for the FS operator:
\begin{equation}
  \delta V_\text{nuc}(r) = \frac{dV_\text{nuc}(r,R_N)}{dR_N} \delta \langle R_N \rangle,
\label{VFS}
\end{equation}
where $R_N$ is the nuclear charge radius. We assume a Fermi distribution for the
nuclear charge. The derivative $dV_\text{nuc}(r,R_N)/dR_N$  is calculated numerically.

The change in the nuclear potential $\delta V_\text{nuc}(r)$ is a small perturbation which
suggests that perturbation theory is probably a proper tool for calculations.
However, the convergence of perturbation theory in the residual Coulomb interaction
is extremely poor. In Table~\ref{tab1}, we present the dependence of 
the FS constants for the $6s$ and $6p$ states of Cs on the number of iterations 
(order of perturbation theory) of the linearized Hartree-Fock equations for an
atom in external field. Here external field is produced by the change of nuclear
radius and corresponding equations are equivalent to the random phase approximation 
(RPA) (see, e.g.~\cite{dzuba87}).
The convergence for the $6s$ state is seen to be very poor while perturbation
theory completely fails for the $6p$ states, inasmuch as one needs more than 
ten iterations to get stable results. 

\begin{table}
\caption{\label{tab1} RPA iterations for the field shift constants $F$ (MHz/fm$^2$) in Cs.}
\begin{ruledtabular}
\begin{tabular}{rccc}
   &     $6s$     &   $6p_{1/2}$   &    $6p_{3/2}$ \\
\hline
 0 &  -1270.24    &    -15.551     &      -.0005   \\
 1 &  -1191.91    &     82.136     &      95.749   \\
 2 &  -1486.35    &      6.227     &      27.535   \\
 3 &  -1341.68    &     52.758     &      71.219   \\
 4 &  -1436.05    &     25.459     &      46.265   \\
 5 &  -1388.48    &     39.612     &      59.382   \\
 6 &  -1413.91    &     32.157     &      52.532   \\
 7 &  -1400.94    &     35.973     &      56.055   \\
 8 &  -1407.65    &     34.003     &      54.241   \\
 9 &  -1404.21    &     35.013     &      55.172   \\
10 &  -1405.98    &     34.494     &      54.694   \\
\end{tabular}
\end{ruledtabular}
\end{table}

Therefore, instead of using perturbation theory, we use an all-order finite-field approach
similar to our early work \cite{dzuba-is}.
Calculations of the FS are done for the reference isotope $A$ with nuclear potential
$V_\text{nuc}(r,R_N(A))$ replaced by
\begin{equation}
  V_\text{nuc}(r,R_N(A)) + \lambda \delta V_\text{nuc}(r),
\label{VFS1}
\end{equation}
where $\lambda$ is scaling parameter. The value of $\lambda$ is chosen in such a way that the 
corresponding change in the nuclear potential is much larger than the numerical uncertainty of the
calculations but is still sufficiently small for the final energy to
be a linear function of $\lambda$. The FS constant for a particular atomic state $v$
is then found as
\begin{equation}
  F_v = \frac{dE_v(\lambda)}{d\lambda}.
\end{equation}
This approach reduces the calculation of the FS to a calculation of energy. We use two 
different techniques to calculate energies. One is the all-order correlation potential
method \cite{dzuba87,dzuba89} 
(also called perturbation theory in the screened Coulomb interaction).
Another is the singles-doubles linearized coupled cluster method 
combined with many-body perturbation theory (MBPT) to account for missing 
third-order diagrams (SD+E3).

\begin{table}
\caption{\label{tab2}Field shift constants $F$ (MHz/fm$^2$) for Cs and Fr 
in different approximations.}
\begin{ruledtabular}
\begin{tabular}{lccccccc}
 & \multicolumn{4}{c}{Cs} & \multicolumn{3}{c}{Fr} \\
           &  \multicolumn{1}{c}{$6s$}  &  \multicolumn{1}{c}{$6p_{1/2}$} 
      &  \multicolumn{1}{c}{$6p_{3/2}$} & \multicolumn{1}{c}{$7p_{1/2}$} 
      &  \multicolumn{1}{c}{$7s$}   
      &  \multicolumn{1}{c}{$7p_{1/2}$} &  \multicolumn{1}{c}{$7p_{3/2}$} \\
\hline
HF   &  -1270 &  -15.7   &  0.0  &   -5.6 & -14111  &  -458  &  0.0 \\
RPA  &  -1405 &   34.6   &   54.9    &  12.4 & -15819  &  -209  &   510   \\
$\langle \hat \Sigma^{(2)} \rangle$   
     &  -2050 &   17.4   &   45.4    &   8.7 & -22358  &  -697  &   313   \\
BO($\hat \Sigma^{(2)}$)          
     &  -2119 &   17.6   &   46.7    &  10.0 & -22447  &  -759  &   301   \\
BO($\hat \Sigma^{\infty}$) 
     &  -1914 &   22.4   &   51.2    &  10.6 & -20463  &  -693  &   303   \\
SD+E3
     &  -1894   & 23.9   &   52.2    &  11.3 & -20188  &  -640  &   361   \\

Other & -2069\footnotemark[1] & 42.78\footnotemark[1] & 70.53\footnotemark[1] & 
15.17\footnotemark[1] & -20782\footnotemark[2] & 
-696\footnotemark[2] & 245\footnotemark[2] \\
\end{tabular}
\end{ruledtabular}
\noindent \footnotetext[1]{Hartley and M{\aa}rtensson-Pendrill
\cite{ann-cs-tl}}
\noindent \footnotetext[2]{M{\aa}rtensson-Pendrill \cite{ann-fr}}
\end{table}

We present the results in Table~\ref{tab2} in order of improving approximations. 
 The first line (HF) gives the average value of the
$\delta V_\text{nuc}(r)$ over Hartree-Fock wave functions. 
The second line (RPA) gives the result
of Hartree-Fock iterations with the potential given by Eq.~(\ref{VFS1}). The result of this
calculation (linear in the
scaling parameter $\lambda$) is equivalent to the RPA.
The next line ($\langle \hat \Sigma^{(2)} \rangle$) includes second-order
correlation corrections by means of many-body perturbation theory. Since we use
``dressed'' basis states (states calculated in a potential given by the Eq.~(\ref{VFS1}) with the FS
operator included) to calculate these corrections, all third-order terms are included
(first in FS and second in Coulomb interaction) as well as chains of higher-order terms
corresponding to core polarization.
We use the notation $\hat \Sigma$ for the correlation 
correction operator (correlation potential). Details of the use of $\hat \Sigma$ in 
atomic calculations can be found elsewhere \cite{dzuba87}.  The line labeled 
BO($\hat \Sigma^{(2)}$)
presents results obtained by including the operator $\hat \Sigma^{(2)}$ in the
Hartree-Fock equations for the valence electron and calculating Brueckner orbitals (BO)
and the corresponding energies. These results differ from those in the previous line by higher-order
contributions in $\hat \Sigma$  ($\hat \Sigma^2,\hat \Sigma^3$, etc.).
Finally, the line BO($\hat \Sigma^{\infty}$) presents results obtained with the all-order
$\hat \Sigma$, in which screening of Coulomb interaction and hole-particle interactions
are included in $\hat \Sigma^{\infty}$ to all orders (see, e.g. \cite{dzuba89} for details)
and  $\hat \Sigma^{\infty}$ is used to calculate BO's.  These are the most accurate results 
obtained in the all-order correlation potential (CP) method. We compare them with results 
obtained in the SD+E3 approach presented in next line. We refer the reader to Ref.~\cite{NA} 
for the details of the (SD+E3) all-order energy calculation.
For the $6s-6p$ and $6s-7p$ intervals (the only ones
important for the analysis of the experimental results), the two methods agree to about 1\%. 

In the last line of Table~\ref{tab2}, we present the FS calculations of
Hartley and M{\aa}rtensson-Pendrill \cite{ann-cs-tl,ann-fr}. They use an approximation
very similar to that listed on line $\langle \hat \Sigma^{(2)} \rangle$.
Naturally, the results are also very close. The larger discrepancy seen for 
the $6p$ states of Cs is probably due
to larger contributions from {\em structural radiation} diagrams (which are called
{\em internal} diagrams in Ref. \cite{ann-cs-tl}). In our approach, calculation of
the IS is reduced to a calculation of energy and no special treatment for the
structural radiation diagrams is needed. In contrast, Hartley and M\'{a}rtensson-Pendrill
evaluated these diagrams using a procedure based on a modification
of the basis. We believe that our calculations are more accurate since
(a) the finite-field approach ensures that no important diagrams are missed, 
(b) we have very good agreement between two very different methods, and
(c) we have very good agreement with experiment for both energies and hyperfine structures
(see below).

The final result of 
Hartley and M\'{a}rtensson-Pendrill for the $6s$ state of Cs 
(-2000~MHz/fm$^2$)
was obtained by rescaling the {\em ab initio} value using a comparison between
theoretical and experimental hyperfine structure (hfs) constants. Their calculated
value for the hfs of $6s$ is larger than the experimental value by 3\%. Therefore,
they reduced the FS constant by the same ratio. Since we include higher-order 
correlations which bring calculated values of the hfs constants and the energies into very
good agreement with experiment, we can check how well this rescaling works.

Table~\ref{tab3} presents the data for the second- and all-order energies, hfs constants, and FS
constants for Cs and Fr. We see that higher-order correlations reduce the ground state energy 
by 3\%, the hfs constant by 5 to 8\% and FS by 9 to 10\%. 
Therefore, energies cannot be used
for rescaling of the FS constants, and accuracy of extracting the higher-order
correlation correction to the FS by rescaling the second-order correction using
hfs data is between 20 and 40\%. This corresponds to 2 to 4\% accuracy in final 
results for $s$-states. Moreover, no reliable rescaling can be done for $p$-states.

Note that the rescaling of FS is very different from the ``fitting of the energy''
procedure used in some of our works (see, e.g. \cite{ginges}) and in the next
section. In that procedure,
we replace $\hat \Sigma$ in equations for valence BO's by the
rescaled operator $\lambda \hat \Sigma$ with $\lambda$ chosen to fit experimental
energies exactly. New BO's are then used to calculate matrix elements. This
procedure changes energies and matrix elements at different rates usually bringing 
the later into significantly better agreement with experiment.

The comparison of final energies and hfs constants with experiments presented in
Table~\ref{tab3} is a further indication that the accuracy of the present calculations of
FS constants is at the level of 1\%.

\begin{table}
\caption{\label{tab3}
Contributions of the higher-order correlations to the energies, hyperfine constants (hfs), and field 
shift constants (FS) of Cs and Fr.}
 \begin{ruledtabular}  
\begin{tabular}{lrrrrr}
 & \multicolumn{1}{c}{BO($\hat \Sigma^{(2)}$)} 
 & \multicolumn{1}{c}{BO($\hat \Sigma^{\infty}$)} & Ratio & Exp\\
\hline
 & \multicolumn{4}{c}{Cesium energies (cm$^{-1}$)} \\
$6s$       & 32375 & 31470 &  0.97 & 31407\footnotemark[1] \\
$6p_{1/2}$ & 20524 & 20296 &  0.99 & 20229\footnotemark[1] \\
$6p_{3/2}$ & 19926 & 19728 &  0.99 & 19675\footnotemark[1] \\
 & \multicolumn{4}{c}{$^{133}$Cs hfs ($g_I$=0.7377208) (MHz)} \\
$6s$       & 2459  & 2270  &  0.92 & 2298.2\footnotemark[2] \\
$6p_{1/2}$ &  314  &  295  &  0.94 & 291.89\footnotemark[3] \\
$6p_{3/2}$ &  51.8 & 48.7  &  0.94 & 50.275\footnotemark[4] \\
 & \multicolumn{4}{c}{Cesium FS (MHz/fm$^{2}$)} \\
$6s$       & -2119 & -1914 &  0.90 & \\
$6p_{1/2}$ &  17.6 & 22.4  &  1.3 & \\
$6p_{3/2}$ &  46.7 & 51.2 &   1.1 & \\

 & \multicolumn{4}{c}{Francium energies (cm$^{-1}$)} \\
$7s$       & 34089 & 32899 &  0.97 & 32849\footnotemark[5] \\
$7p_{1/2}$ & 20986 & 20711 &  0.99 & 20612\footnotemark[5] \\
$7p_{3/2}$ & 19164 & 18976 &  0.99 & 18925\footnotemark[5] \\
 & \multicolumn{4}{c}{$^{211}$Fr hfs ($g_I$=0.888) (MHz)} \\
$7s$       & 9269  & 8769  &  0.95 & 8713.9(8)\footnotemark[6] \\
$7p_{1/2}$ & 1261  & 1193  &  0.95 & 1142.0(3)\footnotemark[7] \\
$7p_{3/2}$ &  98.3 & 102.4 &  1.04 & 94.9(3)\footnotemark[6]\\
 & \multicolumn{4}{c}{Francium FS (MHz/fm$^{2}$)} \\
$7s$       & -22447 & -20463 &  0.91 & \\
$7p_{1/2}$ &  -759  &   -693 &  0.91 & \\
$7p_{3/2}$ &   301  &    303 &  1.01 & \\
\end{tabular}
\end{ruledtabular}
\noindent \footnotetext[1]{Moore \cite{Moore}}
\noindent \footnotetext[2]{Arimondo {\em et al.} \cite{cshfs}}
\noindent \footnotetext[3]{Rafac and Tanner \cite{Rafac}}
\noindent \footnotetext[4]{Tanner and Weiman \cite{Tanner}}
\noindent \footnotetext[5]{Bauche {\em et al.} \cite{Bauche}, Arnold {\em et al.}
\cite{Arnold}}
\noindent \footnotetext[6]{Ekstr\"{o}m {\em et al.} \cite{Ekstrom}}
\noindent \footnotetext[7]{Grossman {\em et al.} \cite{Grossman}}
\end{table}

\subsection{Specific Mass Shift}

The finite-field approach used in the previous section to calculate the field shift
can also be used to calculate the specific mass shift.
To do so we need to redefine the Coulomb interaction in the following way
\begin{equation}
  \langle ab|e^2/r_{12}|cd \rangle \rightarrow \langle ab|e^2/r_{12}|cd \rangle +
  \lambda \langle ab|\mathbf{p_1 \cdot p_2}|cd \rangle,
\label{pp}
\end{equation}
where $\lambda$ is a scaling parameter and $\mathbf{p}$ is the electron momentum 
(see \cite{dzuba-is} for details).

The substitution (\ref{pp}) can be easily done in the Hartree-Fock approximation
or in perturbation theory
calculations. However, the methods used in the previous section to include 
higher-order correlation corrections in the FS constant are not  applicable here.
Note that for FS calculations only the Hartree-Fock program needs to be modified
to incorporate the change in nuclear potential (\ref{VFS}). By contrast, for 
SMS calculations {\em every} program must be modified. While this is still 
straightforward in
Hartree-Fock and perturbation theory codes, it becomes much more difficult 
in higher orders.
In the correlation potential method, inclusion of higher-order correlations is done
by summing the matrix geometric progression \cite{dzuba89}
\begin{equation}
  \tilde Q = Q + Q\Pi Q + Q\Pi Q \Pi Q + ...
\label{QPQ}
\end{equation}
where $Q$ is the Coulomb interaction ($Q_k(r_1,r_2) = r_<^k/r_>^{k+1}$),
$\Pi$ is polarization operator, and $\tilde Q$ is {\em screened} Coulomb interaction
(see, e.q. \cite{dzuba89} for details).
To include the SMS operator in this summation, one would need to modify Coulomb
interaction in the following way
\begin{equation}
  Q \rightarrow Q + \lambda P,
\label{lP}
\end{equation}
where $P$ is the coordinate representation of the SMS operator, which
would lead to the correct expressions for radial integrals (\ref{pp}) when
integrated over the wave functions. It is clear that 
there is no such representation for the SMS operator.
This problem does not appear in the SD+E3 method, since
everything there is expressed in terms of Coulomb integrals which can be
modified according to Eq.~(\ref{pp}). While such modification is as straightforward as for 
perturbation theory codes, technically it is not an easy task. 
The problem here is not only with the large number of terms which must be 
modified, but also with different symmetry properties of the SMS operator. 
Exchanging indexes $a$ and $c$ (or $b$ and $d$) in Eq.~(\ref{pp})
leaves the Coulomb part of the equation unchanged while the SMS contribution changes sign!
While we stress that it would be extremely useful to have a finite-field program for the SMS,
we must leave this for future work.

In the present work, we use two less sophisticated (and less accurate) approaches.
The first is perturbation theory \cite{is3} and the second is a finite-field approach
in which only the second-order correlation operator $\hat \Sigma^{(2)}$ is used.
 In the perturbation theory calculation, we express the SMS operator 
$P=\sum_{i<j}{\bf p}_i \cdot {\bf p}_j$ as a sum of a
normally ordered one-particle operator $S$ and normally ordered two-particle operator $T$.
We carried out an all-order calculation of the matrix element of S; the calculation of the 
matrix element of $T$ is complete through third-order (first order in the SMS operator
and up to second order in Coulomb interaction). 
The results of the perturbation theory calculation for the SMS for Cs and Fr are presented
in Table~\ref{tab4}. The lowest-order values are given in row $S^{(1)}$, the results of the 
all-order singles-doubles  calculation of the matrix element of $S$ are given in row  $SD$, 
and the second- and third-order matrix elements of $T$ are given in rows labeled $T^{(2)}$ 
and $T^{(3)}$. The total values of the specific  mass shift constants are listed in the 
last row.

\begin{table}
\caption{\label{tab4}Perturbation theory contributions to specific mass isotope shift 
constants (GHZ amu) in Cs and Fr}
 \begin{ruledtabular}  
\begin{tabular}{lrrrrrr}
 & \multicolumn{3}{c}{Cs} & \multicolumn{3}{c}{Fr} \\
           &  \multicolumn{1}{c}{$6s$}  &  \multicolumn{1}{c}{$6p_{1/2}$} 
      &  \multicolumn{1}{c}{$6p_{3/2}$} &  \multicolumn{1}{c}{$7s$}   
      &  \multicolumn{1}{c}{$7p_{1/2}$} &  \multicolumn{1}{c}{$7p_{3/2}$} \\
\hline
$S^{(1)}$  & -781.3 &   -191.4   &   -168.7    & -1359.7 &   -260.0   &   -187.0    \\
$SD$       &  316.5 &    139.9   &    140.3    &   363.7 &    182.7   &    182.4    \\
$T^{(2)}$  &  286.8 &     58.5   &     51.2    &   499.9 &     78.9   &     57.5    \\
$T^{(3)}$  & -136.2 &    -30.1   &    -27.8    &  -296.0 &    -53.7   &    -45.0    \\
Total      & -314.2 &    -23.1   &     -5.0    &  -786.1 &    -53.0   &      7.9    \\
\end{tabular}
\end{ruledtabular}
\end{table}

Table~\ref{tab5} presents results of the finite-field approach.
The line {\em HF}  gives the expectation values of the SMS operator over HF
wave function of the valence electron. It is equivalent to the line $S^{(1)}$ of
Table~\ref{tab4}. The difference in numerical values is due to the fact
that the relativistic form of the momentum operator was used in the PT
calculations ($S^{(1)}$) while the non-relativistic operator was used in the
finite-field calculation. Note that relativistic corrections 
for $s - p$ intervals do not exceed 4\% for Cs and 8\% for Fr.

This is a negligible contribution since the mass shift itself is small
for heavy atoms owing to the huge suppression by the mass factor (see Eq.~(\ref{IS})).
Note, however, that relativistic corrections are probably very important for
highly-charged ions. 

The line {\em RPA} in Table~\ref{tab5} presents results of HF iterations with
the SMS operator included in the HF potential by redefining of the Coulomb
interaction according to Eq.~(\ref{pp}). The line $\Sigma^{(2)}$ includes second-order
correlation corrections. Finally, the line {\em Brueck} presents results for
valence electron Brueckner orbitals calculated with using $\Sigma^{(2)}$.
We have also included two ways of simulating higher-order correlations
to try to further improve the results.
Note that Brueckner orbitals with second order $\hat \Sigma$ considerably
overestimate the correlation correction to the energy. They probably have the same effect
on the SMS. Therefore, we reduce the total correlation correction to the SMS 
in two different ways. 

Firstly, we note that iterations of
$\hat \Sigma^{(2)}$ enhance correlation corrections to the SMS 
by a larger factor than for energies. 
If we use the energy ratio to determine the enhancement instead, the total correlation
correction to the SMS is smaller. The corresponding {\em interpolated}
results are presented in the line {\em Interp} in Table~\ref{tab5}.
Note that the correction is huge. It can even change sign of the SMS constant.
This is very different from the FS constants discussed in previous section.
For the FS constants higher-order correction is small and should be treated
accurately. This is why rescaling works for $s$-states only.

Secondly, we rescaled the operator $\Sigma{(2)}$ while calculating Brueckner
orbitals to fit the experimental energies. Scaling factors are
$\lambda (6s) = 0.802, \lambda (6p) = 0.85$ for Cs and
$\lambda (7s) = 0.786, \lambda (7p) = 0.85$ for Fr.
This procedure also reduces the correlation correction to the SMS.
Corresponding results are presented in line {\em Fit} of Table~\ref{tab5}.
It is interesting that the two procedures give close results.

\begin{table*}
\caption{\label{tab5}SMS constants for Cs and Fr in different approximations
(GHZ amu).}
\begin{ruledtabular}
\begin{tabular}{lrrrrrrrr}
 & \multicolumn{5}{c}{Cs} & \multicolumn{3}{c}{Fr} \\
           &  \multicolumn{1}{c}{$6s$}  &  \multicolumn{1}{c}{$6p_{1/2}$} 
      &  \multicolumn{1}{c}{$6p_{3/2}$} &  \multicolumn{1}{c}{$7p_{1/2}$}   
      &  \multicolumn{1}{c}{$7p_{3/2}$} &  \multicolumn{1}{c}{$7s$}   
      &  \multicolumn{1}{c}{$7p_{1/2}$} &  \multicolumn{1}{c}{$7p_{3/2}$} \\
\hline
{\em HF} & -773.18 & -208.70 & -170.40 & -74.68 & -61.44 & -1330.48 & -317.79 & -190.87 \\
{\em RPA}& -355.55 &  -76.96 &  -40.31 & -27.59 & -14.56 &  -666.71 & -127.90 &  -10.48 \\
$\Sigma^{(2)}$
       & -133.19 &  -22.83 &   20.66 &  -5.63 &   8.81 &  -334.15 &  -85.24 &   60.08 \\
{\em Brueck} &    6.75 &    6.85 &   47.33 &   0.95 &  14.31 &  -110.29 &  -45.11 &   89.86 \\
{\em Interp} &  -99.84 &  -14.39 &   29.38 &  -4.31 &  10.21 &  -288.26 &  -77.90 &   69.96 \\
{\em Fit}    &  -89.31 &  -10.26 &   30.17 &  -2.94 &  10.47 &  -311.96 &  -63.83 &   70.35 \\
{\em PT}\footnotemark[1]
       & -314.2 &    -23.1 &   -5.0  &        &        &   -786.1 &  -53.0   &   7.9  \\
Other  &  -23.5\footnotemark[2] & -36.6\footnotemark[2] & 9.2\footnotemark[2] & &
    &  -570\footnotemark[3] &  -154\footnotemark[3] & -18\footnotemark[3] \\
\end{tabular}
\end{ruledtabular}
\noindent \footnotetext[1]{From Table~\ref{tab4}}
\noindent \footnotetext[2]{Hartley and M{\aa}rtensson-Pendrill
\cite{ann-cs-tl}}
\noindent \footnotetext[3]{M{\aa}rtensson-Pendrill \cite{ann-fr}}
\end{table*}

Comparison of results in Table~\ref{tab4} and Table~\ref{tab5} reveals
the poor convergence of perturbation theory and the 
significant difference in
final results between the two calculation methods. Note that the
two methods are
equivalent at second order in the Coulomb interaction. The difference comes
from higher orders. Perturbation theory calculations use the SMS
matrix elements which are just expectation values of the SMS
operator over HF wave functions (HF matrix elements).
By contrast, the finite-field approach corresponds to including
``dressed'' SMS matrix elements in which certain chains
of Coulomb diagrams are included to all orders by iterating the HF
equations (RPA matrix elements). By comparing lines {\em HF} and
{\em RPA} of Table~\ref{tab5} one can see that this indeed must
lead to large differences in final results.
Note that the Brueckner-orbital calculations
are also in better agreement with the calculations of Hartley
and M{\aa}rtensson-Pendrill \cite{ann-cs-tl,ann-fr}.

\begin{table}
\caption{\label{tab6}Field shift constants for K and Rb (MHz/fm$^2$)}
\begin{ruledtabular}
\begin{tabular}{cccc}
K~$4s$ &K~$4p_{1/2}$ &Rb~$5s$ &Rb~$5p_{3/2}$ \\
\hline
-104.20 & 4.04 & -551.85 & 15.60 \\
\end{tabular}
\end{ruledtabular}
\end{table}     

To check whether Brueckner orbitals really give better results
than PT for the SMS, we have performed calculations for potassium and rubidium. 
For these atoms, ``experimental'' values of the SMS can be inferred
by subtracting the NMS and FS from known experimental
values of the isotope shifts. The NMS is given by Eq.~(\ref{NMS}) while the field shift
is calculated as $F \delta \langle r^2 \rangle$. The field shift constant $F$ can
be calculated to high precision as described in the previous 
section. The corresponding values are presented in Table~\ref{tab6}.
Values of $\delta \langle r^2 \rangle$ between the most abundant isotopes
are also known from experimental studies \cite{Angeli}.
Table~\ref{tab7} presents the extraction of the ``experimental'' SMS
between $^{39}$K and $^{41}$K for the $4s-4p_{1/2}$ transition and
between $^{85}$Rb and $^{87}$Rb for the $5s-5p_{3/2}$ transition.
In Table~\ref{tab8} we compare these experimental results with calculations. 
One can see that interpolated, fitted, and unfitted Brueckner 
orbital results are closest to experiment while the perturbation theory
results are much farther away. It is natural, therefore, to take the
average value of the most accurate results ({\em Brueck, Interp} and 
{\em Fit}) as a final central point of 
the calculations, while using differences between these results as an estimate 
of numerical uncertainty. 

\begin{table}
\caption{\label{tab7} Extracting an ``experimental'' SMS for K and Rb.}
\begin{ruledtabular}
\begin{tabular}{ccccc}
 IS(exp) & $\delta \langle r^2 \rangle$\footnotemark[1] & FS & NMS & SMS \\
 (MHz)   & (fm$^2$) & (MHz) & (MHz) & (MHz) \\
\hline
\multicolumn{5}{c}{$^{39-41}$K~~$4s - 4p_{1/2}$} \\
235.25(75)\footnotemark[2] & 0.105 & -11.33 & 267.11 & -20.53(75) \\
\multicolumn{5}{c}{$^{85-87}$Rb~$5s - 5p_{3/2}$} \\
 77.992(20)\footnotemark[3] & -0.042 & 23.83 &  57.00 & -2.84(2) \\
\end{tabular}
\end{ruledtabular}
\noindent \footnotetext[1]{Angeli \cite{Angeli}}
\noindent \footnotetext[2]{Touchard {\em et al.} \cite{Touchard}}
\noindent \footnotetext[3]{Banerjee {\em et al.} \cite{Das}}
\end{table}

\begin{table}
\caption{\label{tab8}Finite-field SMS constants for K and Rb in different approximations
(GHZ amu).}
\begin{ruledtabular}
\begin{tabular}{lrrrrrr}
 & \multicolumn{3}{c}{K} & \multicolumn{3}{c}{Rb} \\
           &  \multicolumn{1}{c}{$4s$}  &  \multicolumn{1}{c}{$4p_{1/2}$} 
      &  \multicolumn{1}{c}{$\Delta$}   &  \multicolumn{1}{c}{$5s$}   
      &  \multicolumn{1}{c}{$5p_{3/2}$} &  \multicolumn{1}{c}{$\Delta$} \\
\hline
{\em HF}     &  -387.27  &  -120.56  &   -266.71  &  -587.21  & -144.92  & -442.29 \\
{\em RPA}    &  -193.62  &   -61.51  &   -132.11  &  -263.55  &  -45.23  & -218.32 \\    
$\Sigma^{(2)}$
       &   -51.71  &   -25.95  &    -25.76  &   -69.46  &    2.45  &  -71.91 \\
{\em Brueck} &    -4.36  &   -15.00  &     10.64  &    14.73  &   19.25  &   -4.52 \\
{\em Interp}  &   -37.52  &   -22.39  &    -15.13  &   -46.17  &    7.69  &  -53.86 \\
{\em Fit}    &   -29.08  &   -19.07  &    -10.01  &   -31.81  &   11.85  &  -43.66 \\
{\em PT}\footnotemark[1]
       &   -74.7   &   -24.7   &    -50.0   &	-163.3  &  -11.8   &  -151.5  \\
Exp    &           &           &    -16.4   &           &          &  -10.5  \\
\end{tabular}
\end{ruledtabular}
\noindent \footnotetext[1]{Safronova and Johnson \cite{is3}}
\end{table}

\section{Results and Discussion}
\label{results}

Final values for the mass and field shift constants are presented in Table~\ref{shift}.
These are the BO ($\hat \Sigma^{\infty}$) results for FS (see Table~\ref{tab2}) and
the average of {\em Brueck, Interp} and {\em Fit} results for SMS (see Table~\ref{tab5}).
We use these values to analyze experimental data and to extract the change in nuclear
charge radius for a range of isotopes of cesium and francium. The results are presented 
in Tables \ref{csrn} and \ref{frrn}. We present two uncertainties for the 
$\delta \langle r^2 \rangle$ for Cs.
The first one is experimental and the second one is theoretical. The theoretical uncertainty 
is dominated by the uncertainty in the SMS constant. 
Strong cancellation between the field shift and the
normal mass shift makes these results sensitive to the SMS. However, the very poor convergence of
perturbation theory make it difficult to predict the SMS to high accuracy. We stress once more
the need for accurate all-order techniques  to calculate the SMS.

\begin{table}
\caption{\label{shift}
Final values of the IS shift constants for Cs and Fr used for the analysis
of the experimental data}
\begin{ruledtabular}
\begin{tabular}{ccrrr}
\multicolumn{2}{c}{Transition} & \multicolumn{1}{c}{$K_\text{SMS}$}
& \multicolumn{1}{c}{$K_\text{NMS}$} & \multicolumn{1}{c}{$F$} \\
 & & GHz amu & GHz amu & MHz/fm$^2$ \\
\hline 
Cs & $6s - 6p_{3/2}$ & -96(56) &  192.9 &  -1965(20) \\
Cs & $6s - 7p_{1/2}$ & -59(65) &  357.9 &  -1925(20) \\
Fr & $7s - 7p_{3/2}$ & -314(113) &  229.0 & -20766(208) \\
\end{tabular}
\end{ruledtabular}
\end{table}

It has been long known (see, e.g. \cite{huber78}) that the change of nuclear radius
along the chain of Cs isotopes is slower than  expected from the formula
$R_N = 1.1 A^{1/3}$. One possible explanation for this fact is that neutrons and 
protons have different distributions. According to data deduced from 
antiprotonic atoms \cite{klos01}, the radius of the neutron distribution for $^{133}$Cs
is by $0.16 \pm 0.06$~fm larger than radius of the proton distribution. It is interesting
to note that under certain assumptions, the isotope shift data is in very close agreement with
the antiprotonic atom data. These assumptions are: (a) the $A^{1/3}$ law is still valid 
but for the total nuclear radius (including the ``neutron skin'') not just charge
radius; (b) the neutron and proton distributions are very close for the neutron poor
nuclei. This is one more argument in favor of the ``neutron skin'' correction
to the PNC in Cs \cite{derevianko}.

\begin{table}
\caption{\label{csrn}
Isotope shift (MHz) and change of nuclear radius (fm$^2$) between $^{133}$Cs and other 
cesium isotopes}
\begin{ruledtabular}
\begin{tabular}{rrrrrrr}
 $A$ & \multicolumn{1}{c}{$6s-$}  &  \multicolumn{1}{c}{SMS} & \multicolumn{1}{c}{NMS} &   
\multicolumn{1}{c}{FS} &  \multicolumn{1}{c}{Exp IS\footnotemark[1]} &
\multicolumn{1}{c}{$\delta \langle r^2 \rangle$} \\
\hline
123 &  $7p_{1/2}$ &  35.88   &  -218.82  &  441.94  &   259(12)   & -0.230(10)(36) \\
124 &  $7p_{1/2}$ &  32.03   &  -195.35  &  424.31  &   261(6)    & -0.220(5)(30)  \\
125 &  $7p_{1/2}$ &  28.25   &  -172.25  &  296.01  &   152(11)   & -0.154(11)(32)  \\
126 &  $7p_{1/2}$ &  24.52   &  -149.53  &  333.01  &   208(7)    & -0.173(6)(23)  \\
127 &  $7p_{1/2}$ &  20.85   &  -127.16  &  200.30  &    94(13)   & -0.104(14)(26)  \\
128 &  $7p_{1/2}$ &  17.24   &  -105.14  &  242.90  &   155(6)    & -0.126(5)(16)  \\
129 &  $7p_{1/2}$ &  13.69   &   -83.46  &  122.77  &    53(9)    & -0.064(11)(18)  \\
		    	       		   	       		   
130 &  $7p_{1/2}$ &  10.19  &   -62.11  &   107.93  &    56(8)    & -0.056(8)(11)  \\
130 &  $6p_{3/2}$ &  16.73  &   -33.48  &    90.75  &  74.0(2.2)  & -0.046(1)(5)  \\
		    	       		   	       		   
131 &  $7p_{1/2}$ &   6.74  &   -41.09  &    25.35  &    -9(6)    & -0.013(8)(11)  \\
131 &  $6p_{3/2}$ &  11.07  &   -22.15  &    21.48  &  10.4(1.6)  & -0.011(2)(7)  \\
		    	       		   	       		   
132 &  $7p_{1/2}$ &   3.34  &   -20.39  &    77.05  &    60(15)   & -0.040(10)(3)  \\
132 &  $6p_{3/2}$ &   5.49  &   -10.99  &    79.80  &  74.3(1.3)  & -0.041(7)(2)  \\
		    	       		   	       		   
134 &  $6p_{3/2}$ &    -5.41  &    10.83  &    27.68  &  33.1(2.5)  & -0.014(1)(1)  \\
135 &  $6p_{3/2}$ &   -10.74  &    21.49  &   -47.15  & -36.4(2.0)  &  0.024(1)(2)  \\
		    	       		   	       		   
137 &  $7p_{1/2}$ &   -12.89  &    78.58  &  -169.70  &   -104(6)   &  0.088(5)(12)  \\
137 &  $6p_{3/2}$ &   -21.17  &    42.36  &  -168.59  & -147.4(2.5) &  0.086(1)(7)  \\
\end{tabular}
\end{ruledtabular}
\noindent \footnotetext[1]{Huber {\em et al.} \cite{huber78}, Thibault {\em et al.}
\cite{thibault81}}
\end{table}

Table~\ref{frrn} presents IS data for Fr. The experimental accuracy for the IS
is very high. Therefore, only the theoretical error is presented for the 
$\delta \langle r^2 \rangle$.
The SMS is less than 1\% of the total
IS for Fr. Also, there is a strong cancellation between the SMS and the NMS, which makes the 
extracted values of the change in nuclear charge radius insensitive to the SMS.
The theoretical uncertainty is therefore quite low and at the level of 1\% which is
mostly the uncertainty in the FS constant. Our results for 
$\delta \langle r^2 \rangle$ differ from those obtained in \cite{ann-fr}
by about 2.4\%.  This is due to the difference in the FS constant in the 
$7s - 7p_{3/2}$ transition. Our value is -20.8 GHz~amu while a value of 
-21.0 GHz~amu is used in \cite{ann-fr}. Since 2\% uncertainty is estimated in
\cite{ann-fr}, 
and we believe that our accuracy is 1\%, we can say that the results are in
good agreement. 

\begin{table}
\caption{\label{frrn}
Isotope shift (MHz) and change of nuclear radius (fm$^2$) between $^{212}$Fr and other
francium isotopes}
\begin{ruledtabular}
\begin{tabular}{rrrrrr}
 $A$  &  \multicolumn{1}{c}{SMS} & \multicolumn{1}{c}{NMS} &   
\multicolumn{1}{c}{FS} &  \multicolumn{1}{c}{Exp IS\footnotemark[1]} &
\multicolumn{1}{c}{$\delta \langle r^2 \rangle$} \\
\hline
207 &      35.73    &  -26.09  &   5229.36   &   5239(4)  & -0.252(3) \\
208 &      28.44    &  -20.77  &   4995.33   &   5003(3)  & -0.241(2) \\
209 &      21.23    &  -15.50  &   3127.27   &   3133(2)  & -0.151(2) \\
210 &      14.09    &  -10.29  &   2599.20   &   2603(1)  & -0.125(1) \\
211 &       7.01    &   -5.12  &    899.11   &    901(3)  & -0.0439(4) \\
213 &      -6.94    &    5.07  &  -1639.13   &  -1641(2)  &  0.0790(8) \\
220 &     -53.78    &   39.28  & -20792.29   & -20806.8(0.5)  &  1.001(10) \\
221 &     -60.23    &   43.99  & -23553.76   & -23570(2)  &  1.134(11) \\
222 &     -66.62    &   48.66  & -26244.03   & -26262(3)  &  1.264(13) \\
223 &     -72.96    &   53.28  & -27902.32   & -27922(2)  &  1.344(14) \\
224 &     -79.24    &   57.87  & -30869.63   & -30891(1)  &  1.487(15) \\
225 &     -85.46    &   62.41  & -32273.95   & -32297(1)  &  1.554(16) \\
226 &     -91.62    &   66.91  & -34376.29   & -34401(1)  &  1.655(17) \\
227 &     -97.74    &   71.38  & -38325.64   & -38352(2)  &  1.846(19) \\
228 &    -103.79    &   75.80  & -40049.01   & -40077(5)  &  1.929(20) \\
\end{tabular}
\end{ruledtabular}
\noindent \footnotetext[1]{Coc {\em et al.} \cite{coc1,coc2}}
\end{table}

Since nuclear radii change quite significantly along the chain of francium isotopes,
it is important to check how PNC matrix elements are affected by this change. 
We have conducted  a numerical test for the change of the matrix element
$W(R_N) \equiv \langle 7s |H^\text{PNC}| 7p_{1/2} \rangle$  with the
change of nuclear radius, assuming that proton and neutron distributions remain the
same. The  numerical results can be presented in a form
\begin{equation}
  \frac{W(R_N)}{W(R_0)} = 1 - 0.21(\frac{R_N}{R_0} -1 ),
\end{equation}
where $R_0$ is the nuclear radius of a reference isotope. Note that total change between
the lightest and the heaviest isotopes in Table~\ref{frrn} is almost 1\%.
This is a significant change and should be taken into account in any future
analysis of the PNC in the chain of Fr isotopes.

Let us now discuss the role that isotope shift calculations may play in the 
study of PNC in atoms.
As it is clear from the results presented above, for atoms in the middle of the
periodic table (Cs, Ba, etc.) the information that can be extracted from the 
IS calculations and used for the PNC analysis is limited by the accuracy of
the SMS calculations. Unless adequate methods are developed to significantly
improve the accuracy of such calculations it is unlikely that the IS data 
will provide information of any practical use for PNC analysis. 

The situation is very different for heavier atoms, such as Fr, Tl, etc. 
The mass shift is small and the corresponding uncertainty can be reduced to an acceptable 
level. On the other hand, calculations of the FS are much easier and can be done with
accuracy of about 1\% or possibly better. There are several possibilities
arising from this fact. To use the IS data to test the electron wave function one
needs to know the value  $\delta \langle r^2 \rangle$  from an
independent source. Such data can be obtained from electron scattering, analysis of 
X-rays from muonic atoms, etc. (see, e.g. \cite{Angeli}).
However, the accuracy of that data is often insufficient for the PNC purposes. 
One can do a consistency test instead. If the isotope shift is known 
for several different
transitions, one can check whether comparison of the theory with experiment
leads to the same value of $\delta \langle r^2 \rangle$ for all transitions.

There is another possibility for many-electron atoms. If the IS is known for an ion 
with simple electronic structure (one electron above closed shells), then
calculations of the IS for this ion can be used to extract the value
of $\delta \langle r^2 \rangle$. Because of simple electronic structure,
the calculations are relatively simple and can be done very accurately.
Then this value of $\delta \langle r^2 \rangle$ can be used in the 
IS and PNC analysis for a neutral atom.

\section{Conclusion}

We have developed methods to calculate the isotope shift for many-electron 
atoms with one external electron above closed shells. While methods
for the field shift seem to be adequate and capable of producing results at
the 1\% or better level of accuracy, methods for the SMS need
further consideration. It would be useful to have an all-order technique
similar to that used in the FS calculations to address the problem
of the very poor convergence of perturbation theory. We use our calculations
for cesium and francium to extract the change in nuclear charge radius for 
chains of isotopes in both atoms. We have demonstrated that, at least for
heavy atoms, calculations and measurements of the isotope shifts may 
provide important information
for the analysis of the PNC in atoms. 

\begin{acknowledgments}

One of the authors (V.D.) is grateful to the
Physics Department of the University of
Notre Dame and Department of Physics and Astronomy
of University of Delaware for the hospitality and support.
The work of W.R.J. was supported in part by NSF grant
No.\ PHY-0139928.

\end{acknowledgments}

\bibliography{csfr}

\begin{thebibliography}{31}
\expandafter\ifx\csname natexlab\endcsname\relax\def\natexlab#1{#1}\fi
\expandafter\ifx\csname bibnamefont\endcsname\relax
  \def\bibnamefont#1{#1}\fi
\expandafter\ifx\csname bibfnamefont\endcsname\relax
  \def\bibfnamefont#1{#1}\fi
\expandafter\ifx\csname citenamefont\endcsname\relax
  \def\citenamefont#1{#1}\fi
\expandafter\ifx\csname url\endcsname\relax
  \def\url#1{\texttt{#1}}\fi
\expandafter\ifx\csname urlprefix\endcsname\relax\def\urlprefix{URL }\fi
\providecommand{\bibinfo}[2]{#2}
\providecommand{\eprint}[2][]{\url{#2}}

\bibitem[{\citenamefont{Murphy et~al.}(2001)\citenamefont{Murphy, Webb,
  Flambaum, Churchill, and Prochaska}}]{alpha1}
\bibinfo{author}{\bibfnamefont{M.~T.} \bibnamefont{Murphy}},
  \bibinfo{author}{\bibfnamefont{J.~K.} \bibnamefont{Webb}},
  \bibinfo{author}{\bibfnamefont{V.~V.} \bibnamefont{Flambaum}},
  \bibinfo{author}{\bibfnamefont{C.~W.} \bibnamefont{Churchill}},
  \bibnamefont{and} \bibinfo{author}{\bibfnamefont{J.~X.}
  \bibnamefont{Prochaska}}, \bibinfo{journal}{Mon. Not. R. Astron. Soc.}
  \textbf{\bibinfo{volume}{327}}, \bibinfo{pages}{1223} (\bibinfo{year}{2001}).

\bibitem[{\citenamefont{Murphy et~al.}(2003{\natexlab{a}})\citenamefont{Murphy,
  Webb, Flambaum, and Curran}}]{alpha2}
\bibinfo{author}{\bibfnamefont{M.~T.} \bibnamefont{Murphy}},
  \bibinfo{author}{\bibfnamefont{J.~K.} \bibnamefont{Webb}},
  \bibinfo{author}{\bibfnamefont{V.~V.} \bibnamefont{Flambaum}},
  \bibnamefont{and} \bibinfo{author}{\bibfnamefont{S.~J.}
  \bibnamefont{Curran}}, \bibinfo{journal}{Asrophys. Space Sci.}
  \textbf{\bibinfo{volume}{283}}, \bibinfo{pages}{577}
  (\bibinfo{year}{2003}{\natexlab{a}}).

\bibitem[{\citenamefont{Murphy et~al.}(2003{\natexlab{b}})\citenamefont{Murphy,
  Webb, and Flambaum}}]{alpha3}
\bibinfo{author}{\bibfnamefont{M.~T.} \bibnamefont{Murphy}},
  \bibinfo{author}{\bibfnamefont{J.~K.} \bibnamefont{Webb}}, \bibnamefont{and}
  \bibinfo{author}{\bibfnamefont{V.~V.} \bibnamefont{Flambaum}},
  \bibinfo{journal}{Mon. Not. R. Astron. Soc.} \textbf{\bibinfo{volume}{345}},
  \bibinfo{pages}{609} (\bibinfo{year}{2003}{\natexlab{b}}).

\bibitem[{\citenamefont{Ginges and Flambaum}(2004)}]{ginges}
\bibinfo{author}{\bibfnamefont{J.~S.~M.} \bibnamefont{Ginges}}
  \bibnamefont{and} \bibinfo{author}{\bibfnamefont{V.~V.}
  \bibnamefont{Flambaum}}, \bibinfo{journal}{Physics Reports}
  \textbf{\bibinfo{volume}{397}}, \bibinfo{pages}{63} (\bibinfo{year}{2004}).

\bibitem[{\citenamefont{Dzuba et~al.}(1986)\citenamefont{Dzuba, Flambaum, and
  Khriplovich}}]{Khriplovich}
\bibinfo{author}{\bibfnamefont{V.~A.} \bibnamefont{Dzuba}},
  \bibinfo{author}{\bibfnamefont{V.~V.} \bibnamefont{Flambaum}},
  \bibnamefont{and} \bibinfo{author}{\bibfnamefont{I.~B.}
  \bibnamefont{Khriplovich}}, \bibinfo{journal}{Z. Phys. D}
  \textbf{\bibinfo{volume}{1}}, \bibinfo{pages}{243} (\bibinfo{year}{1986}).

\bibitem[{\citenamefont{Fortson et~al.}(1990)\citenamefont{Fortson, Pang, and
  Wilets}}]{Fortson}
\bibinfo{author}{\bibfnamefont{E.~N.} \bibnamefont{Fortson}},
  \bibinfo{author}{\bibfnamefont{Y.}~\bibnamefont{Pang}}, \bibnamefont{and}
  \bibinfo{author}{\bibfnamefont{L.}~\bibnamefont{Wilets}},
  \bibinfo{journal}{Phys. Rev. Lett.} \textbf{\bibinfo{volume}{65}},
  \bibinfo{pages}{2857} (\bibinfo{year}{1990}).

\bibitem[{\citenamefont{Hartley and M{\aa}rtensson-Pendrill}(1991)}]{ann-cs-tl}
\bibinfo{author}{\bibfnamefont{A.~C.} \bibnamefont{Hartley}} \bibnamefont{and}
  \bibinfo{author}{\bibfnamefont{A.-M.} \bibnamefont{M{\aa}rtensson-Pendrill}},
  \bibinfo{journal}{J. Phys. B} \textbf{\bibinfo{volume}{24}},
  \bibinfo{pages}{1193} (\bibinfo{year}{1991}).

\bibitem[{\citenamefont{M{\aa}rtensson-Pendrill}(2000{\natexlab{a}})}]{ann-fr}
\bibinfo{author}{\bibfnamefont{A.-M.} \bibnamefont{M{\aa}rtensson-Pendrill}},
  \bibinfo{journal}{Molecular Physics} \textbf{\bibinfo{volume}{98}},
  \bibinfo{pages}{1201} (\bibinfo{year}{2000}{\natexlab{a}}).

\bibitem[{\citenamefont{M{\aa}rtensson-Pendrill}(2000{\natexlab{b}})}]{ann-pnc}
\bibinfo{author}{\bibfnamefont{A.-M.} \bibnamefont{M{\aa}rtensson-Pendrill}},
  \bibinfo{journal}{Hyperfine Interactions} \textbf{\bibinfo{volume}{127}},
  \bibinfo{pages}{41} (\bibinfo{year}{2000}{\natexlab{b}}).

\bibitem[{\citenamefont{Dzuba et~al.}(1987)\citenamefont{Dzuba, Flambaum,
  Silvestrov, and Sushkov}}]{dzuba87}
\bibinfo{author}{\bibfnamefont{V.~A.} \bibnamefont{Dzuba}},
  \bibinfo{author}{\bibfnamefont{V.~V.} \bibnamefont{Flambaum}},
  \bibinfo{author}{\bibfnamefont{P.~G.} \bibnamefont{Silvestrov}},
  \bibnamefont{and} \bibinfo{author}{\bibfnamefont{O.~P.}
  \bibnamefont{Sushkov}}, \bibinfo{journal}{J. Phys. B}
  \textbf{\bibinfo{volume}{20}}, \bibinfo{pages}{1399} (\bibinfo{year}{1987}).

\bibitem[{\citenamefont{Berengut et~al.}(2003)\citenamefont{Berengut, Dzuba,
  and Flambaum}}]{dzuba-is}
\bibinfo{author}{\bibfnamefont{J.~C.} \bibnamefont{Berengut}},
  \bibinfo{author}{\bibfnamefont{V.~A.} \bibnamefont{Dzuba}}, \bibnamefont{and}
  \bibinfo{author}{\bibfnamefont{V.~V.} \bibnamefont{Flambaum}},
  \bibinfo{journal}{Phys. Rev. A} \textbf{\bibinfo{volume}{68}},
  \bibinfo{pages}{022502} (\bibinfo{year}{2003}).

\bibitem[{\citenamefont{Dzuba et~al.}(1989)\citenamefont{Dzuba, Flambaum, and
  Sushkov}}]{dzuba89}
\bibinfo{author}{\bibfnamefont{V.~A.} \bibnamefont{Dzuba}},
  \bibinfo{author}{\bibfnamefont{V.~V.} \bibnamefont{Flambaum}},
  \bibnamefont{and} \bibinfo{author}{\bibfnamefont{O.~P.}
  \bibnamefont{Sushkov}}, \bibinfo{journal}{Phys. Lett. A}
  \textbf{\bibinfo{volume}{140}}, \bibinfo{pages}{493} (\bibinfo{year}{1989}).

\bibitem[{\citenamefont{Safronova et~al.}(1998)\citenamefont{Safronova,
  Derevianko, and Johnson}}]{NA}
\bibinfo{author}{\bibfnamefont{M.~S.} \bibnamefont{Safronova}},
  \bibinfo{author}{\bibfnamefont{A.}~\bibnamefont{Derevianko}},
  \bibnamefont{and} \bibinfo{author}{\bibfnamefont{W.~R.}
  \bibnamefont{Johnson}}, \bibinfo{journal}{Phys.\ Rev.\ A}
  \textbf{\bibinfo{volume}{58}}, \bibinfo{pages}{1016} (\bibinfo{year}{1998}).

\bibitem[{\citenamefont{Moore}(1971)}]{Moore}
\bibinfo{author}{\bibfnamefont{C.~E.} \bibnamefont{Moore}},
  \emph{\bibinfo{title}{Atomic Energy Levels}}, vol.~\bibinfo{volume}{35} of
  \emph{\bibinfo{series}{Natl.\ Bur.\ Stand.\ Ref.\ Data Ser.}}
  (\bibinfo{address}{U.S.\ GPO, Washington, D.C.}, \bibinfo{year}{1971}).

\bibitem[{\citenamefont{Arimondo et~al.}(1977)\citenamefont{Arimondo, Inguscio,
  and Violino}}]{cshfs}
\bibinfo{author}{\bibfnamefont{E.}~\bibnamefont{Arimondo}},
  \bibinfo{author}{\bibfnamefont{M.}~\bibnamefont{Inguscio}}, \bibnamefont{and}
  \bibinfo{author}{\bibfnamefont{P.}~\bibnamefont{Violino}},
  \bibinfo{journal}{Rev. Mod. Phys.} \textbf{\bibinfo{volume}{49}},
  \bibinfo{pages}{31} (\bibinfo{year}{1977}).

\bibitem[{\citenamefont{Rafac and Tanner}(1997)}]{Rafac}
\bibinfo{author}{\bibfnamefont{R.~J.} \bibnamefont{Rafac}} \bibnamefont{and}
  \bibinfo{author}{\bibfnamefont{C.~E.} \bibnamefont{Tanner}},
  \bibinfo{journal}{Phys. Rev. A} \textbf{\bibinfo{volume}{56}},
  \bibinfo{pages}{1027} (\bibinfo{year}{1997}).

\bibitem[{\citenamefont{Tanner and Wieman}(1988)}]{Tanner}
\bibinfo{author}{\bibfnamefont{C.~E.} \bibnamefont{Tanner}} \bibnamefont{and}
  \bibinfo{author}{\bibfnamefont{C.}~\bibnamefont{Wieman}},
  \bibinfo{journal}{Phys. Rev. A} \textbf{\bibinfo{volume}{38}},
  \bibinfo{pages}{1616} (\bibinfo{year}{1988}).

\bibitem[{\citenamefont{Bauche et~al.}(1986)}]{Bauche}
\bibinfo{author}{\bibfnamefont{J.}~\bibnamefont{Bauche}} \bibnamefont{et~al.},
  \bibinfo{journal}{J. Phys. B} \textbf{\bibinfo{volume}{19}},
  \bibinfo{pages}{L593} (\bibinfo{year}{1986}).

\bibitem[{\citenamefont{Alnold et~al.}(1989)}]{Arnold}
\bibinfo{author}{\bibfnamefont{E.}~\bibnamefont{Alnold}} \bibnamefont{et~al.},
  \bibinfo{journal}{J. Phys. B} \textbf{\bibinfo{volume}{22}},
  \bibinfo{pages}{L391} (\bibinfo{year}{1989}).

\bibitem[{\citenamefont{Ekstr\"{o}m et~al.}(1986)\citenamefont{Ekstr\"{o}m,
  Robertsson, Ros\'{e}n, and the ISOLDE~Collaboration}}]{Ekstrom}
\bibinfo{author}{\bibfnamefont{C.}~\bibnamefont{Ekstr\"{o}m}},
  \bibinfo{author}{\bibfnamefont{L.}~\bibnamefont{Robertsson}},
  \bibinfo{author}{\bibfnamefont{A.}~\bibnamefont{Ros\'{e}n}},
  \bibnamefont{and} \bibinfo{author}{\bibnamefont{the ISOLDE~Collaboration}},
  \bibinfo{journal}{Phys. Scr.} \textbf{\bibinfo{volume}{34}},
  \bibinfo{pages}{624} (\bibinfo{year}{1986}).

\bibitem[{\citenamefont{Grossman et~al.}(1999)\citenamefont{Grossman, Orozco,
  Pearson, Simsarian, Sprouse, and Zhao}}]{Grossman}
\bibinfo{author}{\bibfnamefont{J.~S.} \bibnamefont{Grossman}},
  \bibinfo{author}{\bibfnamefont{L.~A.} \bibnamefont{Orozco}},
  \bibinfo{author}{\bibfnamefont{M.~R.} \bibnamefont{Pearson}},
  \bibinfo{author}{\bibfnamefont{J.~E.} \bibnamefont{Simsarian}},
  \bibinfo{author}{\bibfnamefont{G.~D.} \bibnamefont{Sprouse}},
  \bibnamefont{and} \bibinfo{author}{\bibfnamefont{W.~Z.} \bibnamefont{Zhao}},
  \bibinfo{journal}{Phys. Rev. Lett.} \textbf{\bibinfo{volume}{83}},
  \bibinfo{pages}{935} (\bibinfo{year}{1999}).

\bibitem[{\citenamefont{Safronova and Johnson}(2001)}]{is3}
\bibinfo{author}{\bibfnamefont{M.~S.} \bibnamefont{Safronova}}
  \bibnamefont{and} \bibinfo{author}{\bibfnamefont{W.~R.}
  \bibnamefont{Johnson}}, \bibinfo{journal}{Phys. Rev. A}
  \textbf{\bibinfo{volume}{64}}, \bibinfo{pages}{052501}
  (\bibinfo{year}{2001}).

\bibitem[{\citenamefont{Angeli}(1998)}]{Angeli}
\bibinfo{author}{\bibfnamefont{I.}~\bibnamefont{Angeli}},
  \bibinfo{journal}{Heavy Ion Physics} \textbf{\bibinfo{volume}{8}},
  \bibinfo{pages}{23} (\bibinfo{year}{1998}).

\bibitem[{\citenamefont{Touchard et~al.}(1981)}]{Touchard}
\bibinfo{author}{\bibfnamefont{F.}~\bibnamefont{Touchard}}
  \bibnamefont{et~al.}, \bibinfo{journal}{Phys. Lett.}
  \textbf{\bibinfo{volume}{108B}}, \bibinfo{pages}{169} (\bibinfo{year}{1981}).

\bibitem[{\citenamefont{Banerjee et~al.}(2003)\citenamefont{Banerjee, Das, and
  Natarajan}}]{Das}
\bibinfo{author}{\bibfnamefont{A.}~\bibnamefont{Banerjee}},
  \bibinfo{author}{\bibfnamefont{D.}~\bibnamefont{Das}}, \bibnamefont{and}
  \bibinfo{author}{\bibfnamefont{V.}~\bibnamefont{Natarajan}},
  \bibinfo{journal}{Optics Letters} \textbf{\bibinfo{volume}{28}},
  \bibinfo{pages}{1579} (\bibinfo{year}{2003}).

\bibitem[{\citenamefont{Huber et~al.}(1978)}]{huber78}
\bibinfo{author}{\bibfnamefont{G.}~\bibnamefont{Huber}} \bibnamefont{et~al.},
  \bibinfo{journal}{Phys. Rev. Lett.} \textbf{\bibinfo{volume}{41}},
  \bibinfo{pages}{459} (\bibinfo{year}{1978}).

\bibitem[{\citenamefont{Trzci\'{n}ska et~al.}(2001)}]{klos01}
\bibinfo{author}{\bibfnamefont{A.}~\bibnamefont{Trzci\'{n}ska}}
  \bibnamefont{et~al.}, \bibinfo{journal}{Phys. Rev. Lett.}
  \textbf{\bibinfo{volume}{87}}, \bibinfo{pages}{082501}
  (\bibinfo{year}{2001}).

\bibitem[{\citenamefont{Derevianko}(2002)}]{derevianko}
\bibinfo{author}{\bibfnamefont{A.}~\bibnamefont{Derevianko}},
  \bibinfo{journal}{Phys. Rev. A} \textbf{\bibinfo{volume}{65}},
  \bibinfo{pages}{012106} (\bibinfo{year}{2002}).

\bibitem[{\citenamefont{Thibault et~al.}(1981)}]{thibault81}
\bibinfo{author}{\bibfnamefont{C.}~\bibnamefont{Thibault}}
  \bibnamefont{et~al.}, \bibinfo{journal}{Nucl. Phys. A}
  \textbf{\bibinfo{volume}{367}}, \bibinfo{pages}{1} (\bibinfo{year}{1981}).

\bibitem[{\citenamefont{Coc et~al.}(1985)}]{coc1}
\bibinfo{author}{\bibfnamefont{A.}~\bibnamefont{Coc}} \bibnamefont{et~al.},
  \bibinfo{journal}{Phys. Lett. B} \textbf{\bibinfo{volume}{163}},
  \bibinfo{pages}{66} (\bibinfo{year}{1985}).

\bibitem[{\citenamefont{Coc et~al.}(1987)}]{coc2}
\bibinfo{author}{\bibfnamefont{A.}~\bibnamefont{Coc}} \bibnamefont{et~al.},
  \bibinfo{journal}{Nucl. Phys. A} \textbf{\bibinfo{volume}{468}},
  \bibinfo{pages}{1} (\bibinfo{year}{1987}).

\end{thebibliography}

\end{document}